\documentclass[aps,prb,showpacs,twocolumn,superscriptaddress]{revtex4}

\usepackage{amsmath}
\usepackage{graphicx}

\begin{document}

\title{Structure of glassy lithium sulfate films sputtered in
nitrogen (LISON):\\ Insight from Raman spectroscopy and ab initio calculations}

\author{Christian R. M\"uller}
\affiliation{Institut f\"ur Physik, Technische Universit\"at
  Ilmenau, 98684 Ilmenau, Germany}
\author{Patrik Johansson}
\affiliation{Department of Applied Physics, Chalmers University of
   Technology, 41296 G\"oteborg, Sweden}
\author{Maths Karlsson} 
\affiliation{Department of Applied Physics, Chalmers University of
   Technology, 41296 G\"oteborg, Sweden}
\author{Philipp Maass}
\affiliation{Institut f\"ur Physik, Technische Universit\"at
  Ilmenau, 98684 Ilmenau, Germany}
\email{philipp.maass@tu-ilmenau.de}
\homepage{http://www.tu-ilmenau.de/theophys2}
\author{Aleksandar Matic}
\affiliation{Department of Applied Physics, Chalmers University of
   Technology, 41296 G\"oteborg, Sweden}

\date{\today}

\begin{abstract}
Raman spectra of thin solid electrolyte films obtained by sputtering a
Li$_2$SO$_4$ target in nitrogen plasma are measured and compared to
ab initio electronic structure calculations for clusters composed of 28 atoms. 
Agreement between measured and calculated spectra is obtained when
oxygen atoms are replaced by nitrogen atoms and when the
nitrogen atoms form bonds with each other. This suggests that
the incorporation of nitrogen during the sputtering process leads to
structures in the film, which prevent crystallization
of these thin film salt glasses.
\end{abstract}


\maketitle

\section{Introduction}\label{sec:intro}
Glassy thin film electrolytes are materials of considerable
technological interest. They are used in the design of modern solid
state batteries, electrochemical sensors, supercapacitors, and
electrochromic devices. Most of the materials nowadays are fabricated
using the sol-gel methods, but new possibilities are currently
explored by using sputtering techniques. An example are thin films
produced by sputtering a 0.75Li$_2$O-0.25P$_2$O$_5$ target in a
nitrogen plasma (LIPON, \cite{Yu/etal:1997}), which show ionic
conductivities of about $3\times10^{-6}\Omega^{-1}{\rm cm}^{-1}$. This
material has been successfully used in microbatteries
\cite{Park/etal:1999} and electrochromic systems
.\cite{Gerouki/Goldner:1999} Through the mixed network former effect
the ionic conductivity could be further increased to about
$9\times10^{-6}\Omega^{-1}{\rm cm}^{-1}$ by taking a
0.75Li$_2$O-0.25[0.2SiO$_2$-0.8P$_2$O$_5$] target.\cite{Li/etal:2003}

The sputtering technique is in particular interesting, since it
extends the glass forming range. This has recently been observed in
the ternary Li$_2$O-B$_2$O$_3$-Li$_2$SO$_4$
systems,\cite{Joo/etal:2003} where the sulfate content could be
increased to amounts that would lead to crystallization when preparing
glass by melt quenching. In fact, the borate component can be fully
eliminated by sputtering a Li$_2$SO$_4$ target in a nitrogen
containing plasma, leading to an amorphous material referred to as
LISON.\cite{Joo/etal:2004} With respect to the target material, one
could speak of a ``salt glass''. It is also found that sputtering in
argon plasma does not lead to an amorphous film, indicating that
nitrogen plays a crucial role in the glass forming ability. Indeed, it
was shown that nitrogen is partly incorporated into the material
\cite{Joo/etal:2004} with a sulphur to nitrogen ratio of about 2:1. At
present it is not clear how the structure of these salt glasses is
built up, in particular in which way the incorporation of nitrogen
takes place and how this leads to a stabilization of the amorphous
structure.

The aim of this work is to get insight in the structural arrangements
of this new glass system, how the nitrogen is incorporated in the
structure and how this is related to the increased stability of the
material towards crystallization. To this end we study LISON films by
Raman spectroscopy and compare the experimental vibrational spectra
with predictions from ab initio electronic structure calculations
based on density functional theory. Calculations have been performed
for pure Li$_2$SO$_4$ clusters and nitrogen doped Li$_2$SO$_4$
clusters.

\section{Experimental}\label{sec:raman}

The Raman experiments were performed at room temperature in argon
atmosphere using a micro-Raman setup with an argon-krypton laser tuned
to the 514~nm line as excitation source. With this experimental set-up
we have a depth resolution of 14~$\mu{\rm m}$. The power at the sample
was 21~mW for the depolarized configuration (perpendicular direction
of polarization of the incoming beam with respect to the polarizer in
front of the detector) and 7~mW for polarized configuration. The
integration time of a single measurement was 300~s. The analysis of
the spectrum is based on the median of 27 spectra. Measurements on
different spots of the sample gave the same spectrum, showing that the
film was homogeneous.

The LISON film had a thickness of $4\,\mu{\rm m}$ and was prepared by
sputtering a Li$_2$SO$_4$ target in nitrogen plasma at pressure
1~Pa.\cite{comm:sample} Due to the rather strong Raman response of the
silicon substrate and the small thickness of the film, the measured
spectra are dominated by the silicon signal. In order to obtain the
response of the film, a reference spectrum from a clean silicon
substrate was subtracted. In this procedure it was important to ensure
that the substrate had the same orientation with respect to the
incident polarization, wherefore we made use of the polarization
dependence of the intensity of the 521~cm$^{-1}$ Raman line of
silicon.
  
\section{Ab initio calculations}\label{sec:abinito}

Electronic structure calculations were performed for Li$_2$SO$_4$
clusters with 28 atoms (4 Li$_2$SO$_4$ units). Start configurations
were created based on the conception that the lithium sulfate salt
glass consists of intact SO$_4$ units with lithium ions in between.
Using the GaussView3.0 program,\cite{gaussview3.09:2003} SO$_4$
tetrahedra were constructed with random orientation. Lithium ions were
added under the constraint of minimum interatomic distance of
1.6~\AA.

The starting configurations were geometry optimized using the
Gaussian03 suite of quantum chemistry programs \cite{gaussian03:2004}
(for the theoretical background underlying the methods see
\cite{Cramer:2004}). First, the semi-empirical PM3 method was applied,
followed by a Hartree Fock calculation with the 6-31G* basis set.
These calculations took about 200~CPUh and $\sim200$ iterations.
For the final optimization using the hybrid density functional B3LYP
and the same basis set only a few iterations were needed.

In order to monitor the influence of nitrogen, modified Li$_2$SO$_4$
clusters were generated by replacing oxygen atoms by nitrogen atoms in
the optimized Li$_2$SO$_4$ cluster configurations. These modified
clusters were again optimized by employing Hartree Fock and B3LYP with
the 6-31G* basis set.

{}From the 2nd and 3rd order derivatives of the potential energy
surface (PES) at the minima found for the optimized configurations,
the Raman spectra were calculated. It is clear that the amorphous film
will exhibit a large number of local structures corresponding to
different minima of the PES. By optimizing different starting
configurations it is possible to scan a part of the PES and find
minima, whose associated clusters are good representatives of the
local structures of the LISON film. This is evaluated by comparing
calculated vibrational spectra with the experimental.

\section{Experimental Results}

\begin{figure}[t!]
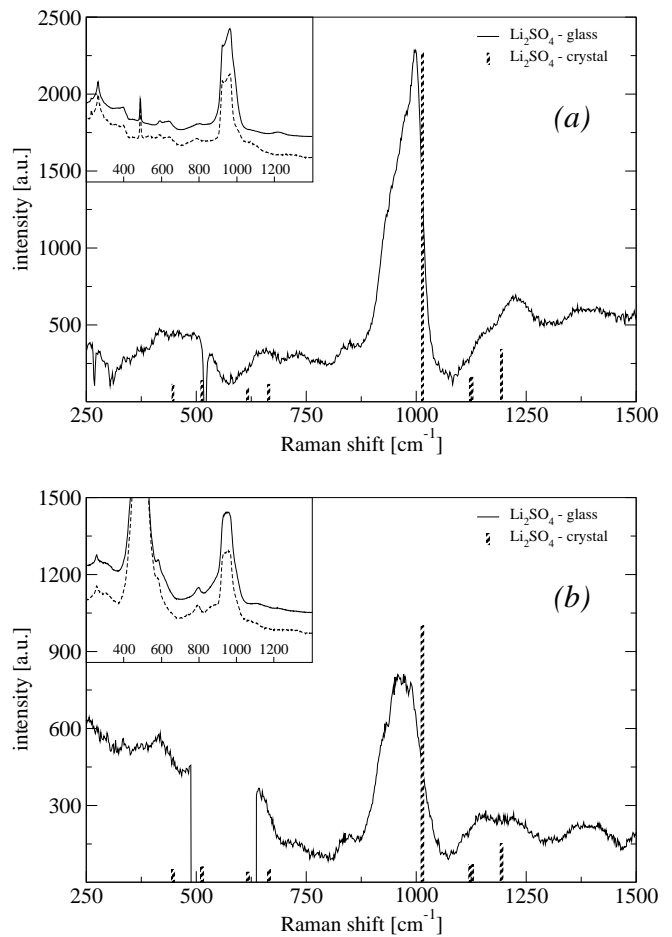

\includegraphics[width=0.48\textwidth,clip=,]{./figures/fig1a.eps}\\[3ex]
\includegraphics[width=0.48\textwidth,clip=,]{./figures/fig1b.eps}
\caption{\label{fig:lison-exp} {\it (a)} Polarized and {\it (b)}
depolarized difference Raman spectra of LISON films on a silicon
substrate. The difference signal is obtained by subtracting the
spectrum of a bare silicon substrate from the spectrum of the sample.
Inset shows the raw data from the sample and the substrate (dashed
lines silicon, solid lines LISON film sample). The spectra have been
offset for clarity.}
\end{figure}

Figure~\ref{fig:lison-exp} shows {\it (a)} the polarized and {\it (b)}
depolarized Raman spectra for the LISON film. The bare silicon (dashed
line) and the sample spectrum (film plus substrate, solid line) are
shown in the insets of Figs.~\ref{fig:lison-exp}a,b. The bare silicon
spectrum was scaled to fit the characteristics between 250~cm$^{-1}$
and 350~cm$^{-1}$. The main feature of the resulting difference
spectra in Figs.~\ref{fig:lison-exp}a,b is a band between
900~cm$^{-1}$ and 1035~cm$^{-1}$. A second weaker band is found at
1225~cm$^{-1}$ with a shoulder at slightly lower frequencies, around
1150~cm$^{-1}$, in the polarized spectrum in
Fig.~\ref{fig:lison-exp}a. In addition two more bands may be discerned
at 415~cm$^{-1}$ and at 645~cm$^{-1}$, in particular in the case of
the polarized spectrum, where the intense silicon mode at
521~cm$^{-1}$ is absent and does not hamper the subtraction procedure.
The intensity gap between 440~cm$^{-1}$ and 500~cm$^{-1}$ in the
depolarized spectrum results from the fact that, in order to resolve
the weak difference signals, the integration time had to be chosen
sufficiently large. As a consequence the intensity of the silicon mode
at 521~cm$^{-1}$ present in the depolarized spectrum (see inset in
Fig.~\ref{fig:lison-exp}b) saturates the detector in this spectral
region.

In order to assign the bands originating from the LISON film in the
difference spectra in Fig.~\ref{fig:lison-exp} we compare the results
to the Raman modes found in crystalline Li$_2$SO$_4$
,\cite{Cazzanelli/Frech:1984} marked as bars in the figure. The Raman
spectrum of the crystal is dominated by the symmetric breathing mode
of SO$_4^{2-}$ at 1014~cm$^{-1}$, which either is very weak or found
at lower frequencies in the glassy film. That this main band covers a
range of frequencies where there are no corresponding modes from the
crystal, suggests that the structure of the amorphous LISON film and
the Li$_2$SO$_4$ crystal are different also on the short range order
scale. The modes at 1123~cm$^{-1}$ and 1127~cm$^{-1}$ in the crystal
(appearing as a single bar in Figs.~\ref{fig:lison-exp}a,b due to
their small separation), as well as the mode at 1194~cm$^{-1}$ seem on
the other hand to be shifted to higher frequencies and can be related
to the bands around 1150~cm$^{-1}$ and around 1225~cm$^{-1}$ in the
experimental spectrum. Similarly, the two bands in the film around
415~cm$^{-1}$ and 645~cm$^{-1}$ have likely their origin in the pair
of modes at 447~cm$^{-1}$, 513~cm$^{-1}$ and the pair of modes at
617~cm$^{-1}$, 665~cm$^{-1}$ in the crystal, respectively.

\section{Theoretical results and comparison with the experiment}
\label{subsec:comparison}

For the comparison of the experimental and calculated spectra we will
exclusively use the polarized Raman spectrum shown in
Fig.~\ref{fig:lison-exp}a, as it is not disturbed by the silicon mode
at 521~cm$^{-1}$. To obtain a smooth spectrum from the calculated
Raman lines, each line $j$ with Raman shift $\omega_j$ and intensity
$I_j$ is replaced by a Gaussian peak function
$I_j\exp[-(\omega-\omega_j)^2/\Delta^2]$ with $\Delta=20\,{\rm
cm}^{-1}$.

In addition, a rescaling of the calculated Raman shifts is in general
necessary to obtain good agreement with the experimental spectra,
depending on the particular material and method used. Commonly used
literature data for these rescaling factors exist for some gas and
organic molecules. They have been determined by adjusting calculated
to measured frequencies for well defined modes. In our case such
literature data are not available and we have to determine the
rescaling factor by other means. To do this we focus on the SO$_4$
breathing mode of the crystal, since one should expect it to be
present also in the clusters and the film (although, possibly, with
reduced intensity). The calculated SO$_4$ breathing mode frequencies
differ slightly from each other due to the different local
surroundings of the SO$_4^{2-}$ anions in the clusters. Taking the
average of them and determining the quotient with the corresponding
crystal mode frequency gives a stretching factor of 1.11. Using this
factor yields theoretical spectra which appear to be shifted to higher
frequencies. An optimal overlap of calculated and measured spectra is
obtained with a slightly reduced factor of 1.09. This minor correction
should not be surprising as some shift of the breathing mode frequency
can be expected when considering the crystal and the film. To
summarize, for the comparison all calculated spectra are stretched by
this factor of 1.09.

\begin{figure}[t!]
\includegraphics[width=0.48\textwidth,clip=,]{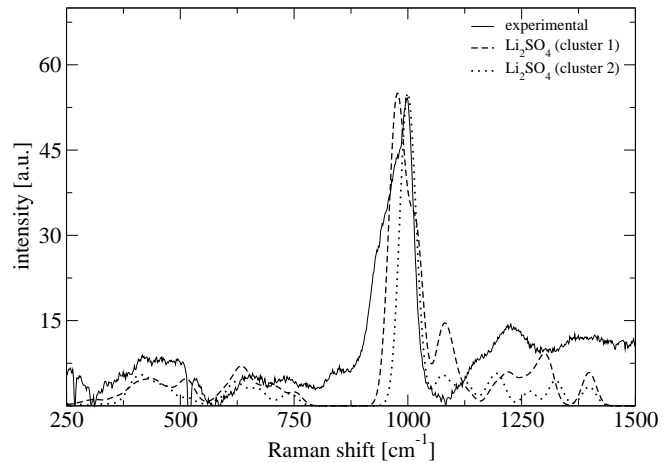}
\caption{\label{fig:lison-comp1} Comparison of two calculated Raman
  spectra for Li$_2$SO$_4$ clusters with the experimental Raman data.}
\end{figure}

\subsection{Li$_2$SO$_4$ clusters}

Figure~\ref{fig:lison-comp1} shows the Raman spectra calculated from
two optimized Li$_2$SO$_4$ cluster configurations (dotted and dashed
lines) in comparison with the measured Raman spectrum (solid line).

As can be seen from the figure, the calculated spectra for the two
different optimized clusters do not deviate much. This can be
interpreted in the way that local arrangements participating in the
dominating vibrations are still determined by the Li$^+$ and
SO$_4^{2-}$ ions and their mutual Coulomb interaction as in the
crystal. As a consequence, the SO$_4^{2-}$ units are rather weakly
coupled.

In comparison to the measured Raman spectrum, the calculated ones
resemble the overall band structure. However, there are two important
differences : {\it (i)} The experimental spectrum shows a broad
shoulder towards lower frequencies from the main band at
1002~cm$^{-1}$ down to 900~cm$^{-1}$. As there exists not a single
vibrational mode between 760~cm$^{-1}$ and 960~cm$^{-1}$, this
shoulder is absent in the calculated spectra. {\it (ii)} In the
calculated spectra there is intensity due to a mode at 1080~cm$^{-1}$,
while almost no intensity is seen around this frequency in the
experimental spectrum. As a first hypothesis deviation {\it (ii)}
might be related to the finite size of the calculated cluster. A
closer inspection of the eigenvector belonging to this mode reveals
that it contains an oxygen atom, which moves in the normal direction
of the surface of the cluster. Accordingly, this mode can be
attributed to an artefact of the limited cluster size.

Deviation {\it (i)} on the other hand might be due to structural
changes when nitrogen is incorporated in the structure, which can be
investigated by comparing experimental spectra to calculated spectra
from the nitrogen containing clusters .

\subsection{Li$_2$SO$_4$-N (LISON) clusters}

\begin{figure}[t!]
\includegraphics[width=0.48\textwidth,clip=,]{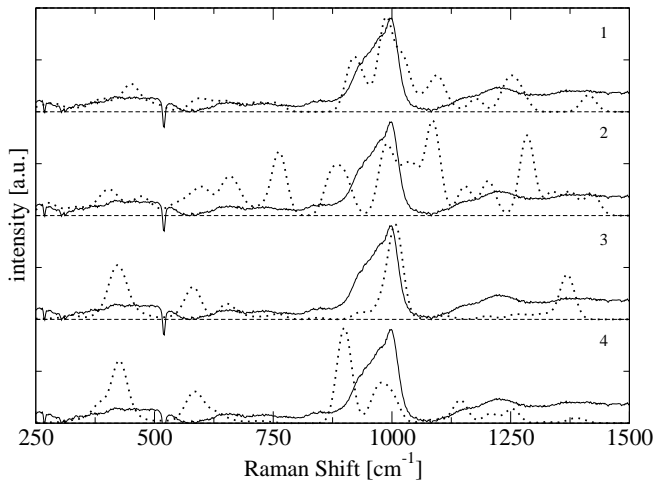}
\caption{\label{fig:lison-comp2} Comparison of four calculated Raman spectra
  (dotted lines) for LISON clusters with the experimental Raman data
  (solid lines).}
\end{figure}

The substitution of two oxygen atoms by nitrogen atoms in a cluster
has a strong effect on the calculated spectra, as can be seen from
Fig.~\ref{fig:lison-comp2}, where spectra of 4 different clusters
(labeled 1-4 in the figure) are compared to the measured Raman
spectrum. This is not surprising as half of the SO$_4^{2-}$ are
replaced by SO$_3$N$^{3-}$. The rather large amount of nitrogen
correspond to results from Rutherford backscattering experiments which
yielded a ratio of about 2:1 of sulfur to nitrogen
\cite{private:yohann} and are also in agreement with previously
reported values.\cite{Joo/etal:2004}

A number of new bands appear in the calculated spectra from the
nitrogen containing clusters. However, while the clusters without
nitrogen lead to very similar spectra, the modified systems show
pronounced differences. In general the spectra 2-4 do not compare well
with the experimental data, while the calculated spectrum 1 is
surprisingly close to the experimental data. In particular, the broad
shoulder of the main band is here well reproduced. The additional band
at 1080~cm$^{-1}$ is still present, but does not relate to nitrogen
incorporation as discussed above. It is interesting to note that the
optimal cluster 1 does not only compare most favorably with the
experimental spectrum, but also has the lowest energy (1.9~eV lower
than the next lowest energy cluster associated with spectrum 3).

\begin{figure}[h!]
\includegraphics[width=0.4\textwidth,clip=,]{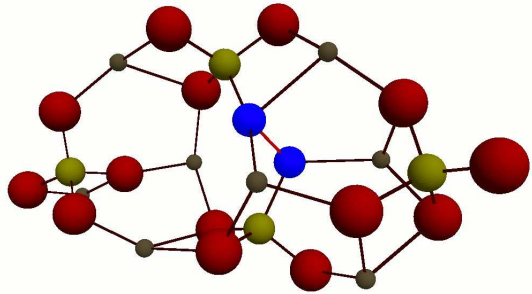}
\caption{\label{fig:cluster} Atomic configuration of the optimal LISON
  cluster (corresponding to spectrum 1 in Fig.~\ref{fig:lison-comp2}).
  Oxygen atoms are marked in red (dark gray, large spheres), sulfur
  atoms in yellow (light gray, mid-size spheres), nitrogen atoms in
  blue (dark grey, mid-size spheres), and Li atoms in silver (grey,
  small spheres).}
\end{figure}

Figure~\ref{fig:cluster} shows the atomic configuration of the optimal
cluster 1. In contrast to the clusters 2-4, it contains an interaction
between two nitrogen atoms in close proximity at a distance of
1.47~\AA, which serves as a type of bridge between the two SO$_3$N
pseudo-tetrahedra. This suggests that the incorporation of nitrogen
triggers the formation of structures, which, by hindering a
reorientation of the anions relative to each other, prevent the
crystallization of the film (or, more precisely, increases the free
energy barrier for crystallization).

\section{Conclusions}\label{sec:conclusions}

By comparing experimental and calculated Raman spectra for LISON
systems, we are able to show that 
a good representation of an
amorphous LISON film is possible using small atomic clusters. Our
investigations reveal that in the LISON structure nitrogen is
replacing oxygen in the SO$_4^{2-}$ tetrahedra. In the configuration
that results in Raman spectra in agreement with the experimental data
we find a nitrogen-nitrogen bridge between two anions. This might be
the underlying reason for the prevention of crystallization of the
LISON material during the sputtering process, as these nitrogen
bridges restrain the reorientation of the anions and therefore
stabilize an amorphous film structure.

\section*{Acknowledgments}
We thank Yohann Hamon and Philippe Vinatier from \'Ecole Nationale
Sup\'erieure de Chimie et Physique de Bordeaux, France, for providing
the LISON samples, and Efstratios Kamitsos from the Theoretical and
Physical Chemistry Institute of the National Hellenic Research
Foundation, Athens, Greece, for valuable discussions. Financial
support by the HI-CONDELEC EU STREP project (NMP3-CT-2005-516975) is
gratefully acknowledged.


\begin{thebibliography}{22}
\bibitem{Cazzanelli/Frech:1984} E.\ Cazzanelli and R.\ Frech,
J.~Chem.~Phys.\ {\bf 79}, 2615 (1983); {\it ibid.} {\bf 81}, 4729
(1984); only the modes belonging to the $A_g$ symmetry are considered
(other representations have slightly shifted modes, but the size of
the corresponding shifts are negligible in comparison with the width
of the peaks obtained for the film).
\end{thebibliography}
\end{document}